\documentclass[epj,final]{svjour}

\usepackage{graphicx}
\usepackage{dcolumn}   
\usepackage{bm}        
\usepackage{amssymb}   
\usepackage{textcomp}
\usepackage{ucs}
\usepackage{multirow}
\usepackage{hyperref}
\usepackage{amsmath}
\usepackage{color}
\usepackage{footnote}
\usepackage{color}
\usepackage{colortbl}

\definecolor{dunkelgrau}{rgb}{0.8,0.8,0.8}
\definecolor{hellgrau}{rgb}{0.95,0.95,0.95}
\definecolor{hellblau}{rgb}{0,1,1}

\begin{document}
\title{Ion-Beam Excitation of Liquid Argon}
\author{M. Hofmann\inst{1,4} \and T. Dandl\inst{2} \and T. Heindl\inst{2} \and A. Neumeier\inst{1} \and L. Oberauer\inst{1} \and W. Potzel\inst{1} \and S. Roth\inst{1} \and S. Sch\"onert\inst{1} \and \\J. Wieser\inst{3} \and A. Ulrich\inst{2}\thanks{\emph{Andreas Ulrich:} andreas.ulrich@ph.tum.de 
} 
}                     
\institute{Technische Universit\"at M\"unchen, Physik-Department E15, James-Franck-Str. 1, D-85748 Garching, Germany \and Technische Universit\"at M\"unchen, Physik-Department E12, James-Franck-Str. 1, D-85748 Garching, Germany \and excitech GmbH, Branterei 33, 26419 Schortens, Germany \and now at: KETEK GmbH, Hofer Str. 3, 81737 M\"unchen, Germany}

\date{Published in Eur. Phys. J. C (2013)}

\abstract{
The scintillation light of liquid argon has been recorded wavelength and time resolved with very good statistics in a wavelength interval ranging from 118\,nm through 970\,nm. Three different ion beams, protons, sulfur ions and gold ions, were used to excite liquid argon. Only minor differences were observed in the wavelength-spectra obtained with the different incident particles. Light emission in the wavelength range of the third excimer continuum was found to be strongly suppressed in the liquid phase. In time-resolved measurements, the time structure of the scintillation light can be directly attributed to wavelength in our studies, as no wavelength shifter has been used.
These measurements confirm that the singlet-to-triplet intensity ratio in the second excimer continuum range is a useful parameter for particle discrimination, which can also be employed in wavelength-integrated measurements as long as the sensitivity of the detector system does not rise steeply for wavelengths longer than 190\,nm. Using our values for the singlet-to-triplet ratio down to low energies deposited a discrimination threshold between incident protons and sulfur ions as low as $\sim$\,2.5\,keV seems possible, which represents the principle limit for the discrimination of these two species in liquid argon.
}           

\maketitle

\section{Introduction}
\label{sec::Introduction}

Liquid rare gases (LRg) are widely used as detector media in various fields like astroparticle physics, high-energy particle physics, or medical diagnostics \cite{noblegasdetectors,curioni,raregasdetectors}. Experiments in astroparticle physics, oftentimes searching for rare events at an energy scale of a few to tens of keV like direct search for Dark Matter in the form of weakly interacting massive particles (WIMPs), search for neutrinoless double beta decay (0$\nu\beta\beta$), or coherent neutrino nucleus scattering (CNNS), in particular, benefit from the outstanding particle detection properties of LRg. The high light yields \cite{doke88} improve photon statistics notably, which is particularly advantageous for low amounts of energy deposited. Besides this, LRg can be scaled easily in quantity allowing their use in large underground detectors. 
In addition, they can be cleaned chemically to a high level of purity, which is important in terms of background: radioactive impurities can be removed from the LRg efficiently, as well as even some of the long-lived radioisotopes of other rare gases like $^{85}$Kr from xenon \cite{bolozdynya}. Also impurities like oxygen or water which alter the light yield \cite{oxygen_in_argon} or light-emission time constants \cite{oxygen_in_argon,heindl,heindl_epl,heindl_jinst} can be removed.

However, mostly for technical reasons not all rare gases turn out to be well-suited for astroparticle physics experiments. The boiling temperatures of helium and neon are far below the temperature of liquid nitrogen (LN2), which makes cooling rather complex and expensive. In addition, their scintillation light is dominantly emitted at wavelengths shorter than 90\,nm \cite{heindl,chepel,xuv_paper,carman,jortner}\footnote{The references quoted all cover the scintillation light emission of neon and/or helium either in the gas or the solid phase. However, as is known from the heavier rare gases, the scintillation spectra of the liquid phase are similar, see e.g. \cite{jortner,cheshnovsky}. Direct spectroscopic data for liquid neon and liquid helium are hard to obtain.}, and no window materials are available which are transparent in this wavelength region. 
Furthermore, the cross section for coherent WIMP scattering scales with the square of the atomic mass number A \cite{chepel}, also favouring the heavier rare gases.
Krypton would, in principle, be a suitable candidate as detector medium, however, it contains non-negligible traces of the radioisotope $^{85}$Kr \cite{raregasdetectors} which beta decays with a Q-value of 687\,keV \cite{firestone}. 
For these reasons most of the currently operational or planned experiments focus on either argon or xenon. An overview of the field can be found in \cite{chepel}. Natural atmospheric argon, however, contains about 1\,$\frac{\text{Bq}}{\text{kg}}$ of $^{39}$Ar \cite{chepel} (269\,yr half life, Q-value: 565\,keV), argon from special sub-surface sources about a factor of 20 less \cite{subsurface}.

This radioactive contaminant in argon in particular, and the high sensitivity which is needed in rare event physics in general requires efficient background rejection, which can partly be achieved by particle discrimination on an event-by-event basis. This can either be carried out by simultaneously recording the scintillation light and the produced charge, as in two-phase TPC detectors, e.g. \cite{ArDM,WARP,darkside}, or by the time-dependence of the scintillation light alone \cite{MiniCLEAN,DEAP}. For the use of commercially available inexpensive photomultipliers the scintillation light, which is predominantly emitted in the vacuum ultraviolet (VUV) region (below 200\,nm; see section \ref{sec::wavelengthstudies}), is usually shifted to UV or visible light by the use of wavelength shifters. Thereby, however, all spectral information of primary light emission is lost, and its time structure is convolved with the scintillation time constant of the wavelength shifter. 

Commonly, particle discrimination by the use of the scintillation light alone relies on the wavelength-integrated intensity ratio of the fast component of the scintillation light to the slow one \cite{MiniCLEAN,DEAP,boulay_psd,peiffer,lippincott}, which is different for particles with different linear energy transfer. The dominant contribution to the fast component of the scintillation light comes hereby from decays of vibrationally relaxed, neutral rare gas excimer molecules in the singlet state\footnote{A detailled explanation of the scintillation light emission processes and the corresponding continua is given in sec. \ref{sec::mechanismus}.}, while the slow components stem from decays of molecules in the triplet state. These two components consequently have (nearly) the same wavelength.
The loss of spectral information of the scintillation light in the measurement, however, might influence this intensity ratio: for example, for the case of neon in the gas phase it has been shown that the third excimer continuum\footnote{An excimer continuum originating from multiply ionized species; again, see sec. \ref{sec::mechanismus}.} only appears for the excitation with heavy ions, but not for the excitation with electrons \cite{xuv_paper}. As light emission in this continuum is also very fast, see sec. \ref{sec::timestudies}, it might be misidentified as part of the fast component of the second excimer continuum.
This can not be disentangled in wavelength-integrated measurements. For this reason we have studied the scintillation light emission of liquid argon (LAr) in a wide wavelength interval (118\,nm - 970\,nm) both wavelength and time resolved with good resolution and very good statistics. All measurements presented in this work were carried out without using a wavelength shifter, therefore, the time structure of the scintillation light emission can directly be attributed to wavelength.

In section \ref{sec::setup} we present the experimental setup which was used to record the scintillation light of liquid argon. In section \ref{sec::results} we show the results of our studies and give, where possible, an interpretation based on the underlying gas kinetic processes. In section \ref{sec::conclusion} we finally conclude on our results and the particle discrimination potential of LAr based on the wavelength and time information of the scintillation light.

\section{Experimental Setup}
\label{sec::setup}

\begin{figure}
	\centering
	\includegraphics[width=\columnwidth]{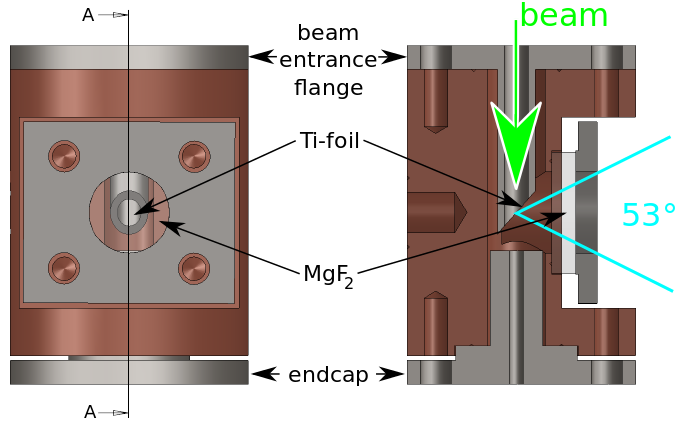}
	\caption{\label{fig::aufbauzelle} \textit{Technical drawing of the target cell. The left side shows the cell from the direction of the light detector; the right picture is a cross-sectional view along plane ''A''. In the experiments the ion beam came from the top (indicated by the arrow; green in color online). The cell was made of copper, and had an outer diameter of 50\,mm and a length of 60\,mm. The inner borehole had a diameter of 12\,mm providing a 5.6\,cm$^3$ volume for the LAr. At one side the cell had a MgF$_2$-window with a thickness of 5\,mm, which was held by a steel plate. Towards the incoming beam the cell was closed with a beam entrance flange which extended into the cell up to the exit window. There, it was inclinedly cut and closed with a titanium foil. On the opposite side the cell was closed with an endcap, which served as a Faraday cup.}}
\end{figure}

For the studies presented in this paper a target cell made of copper was designed, see fig. \ref{fig::aufbauzelle}. It had a length of 60\,mm, an outer diameter of 50\,mm, and an inner borehole with a diameter of 12\,mm. The volume which was filled with LAr during operation was 5.6\,cm$^3$. On one side the cell had a MgF$_2$-window, which allowed the scintillation light to exit the cell. MgF$_2$ is transparent down to $\sim$\,113\,nm \cite{samson}, therefore, the emission spectrum of LAr could be recorded down to the short-wavelength end of the second excimer continuum. The opening angle for the scintillation light was $\sim$\,53$^{\circ}$. Thus, the exit port of the surrounding CF-100 cross piece was fully illuminated and no light was lost by shadowing effects.

Towards the incoming ion beam the cell was closed with a beam entrance flange made from stainless steel with an inner borehole of 5\,mm, extending into the cell up to the MgF$_2$ exit window. At its end this flange was closed by a 1.5\,$\frac{\text{mg}}{\text{cm}^2}$-thick titanium foil, which separated the beam line vacuum from the cell filled with LAr and allowed the ion beam to enter the liquid. As the range of heavy ions in LAr is only a few tens of micrometers most of the scintillation light was produced directly in front of the foil. Therefore, the entrance flange was cut under an angle of 45$^{\circ}$ at its end in such a way that the effective light emitting area seen by the light detector was maximized. The distance the scintillation light had to travel through LAr was minimized to only few millimeters, thereby avoiding absorption effects \cite{neumeier} and effects related to Rayleigh scattering \cite{seidel_ar}.
On the opposite side the cell was closed with a stainless steel endcap, which had an electrically insulated copper plate at its front. This plate served as Faraday cup for the ion beam for an evacuated cell and thereby allowed us finding an optimal beam adjustment.

\begin{figure}
	\centering
	\includegraphics[width=\columnwidth]{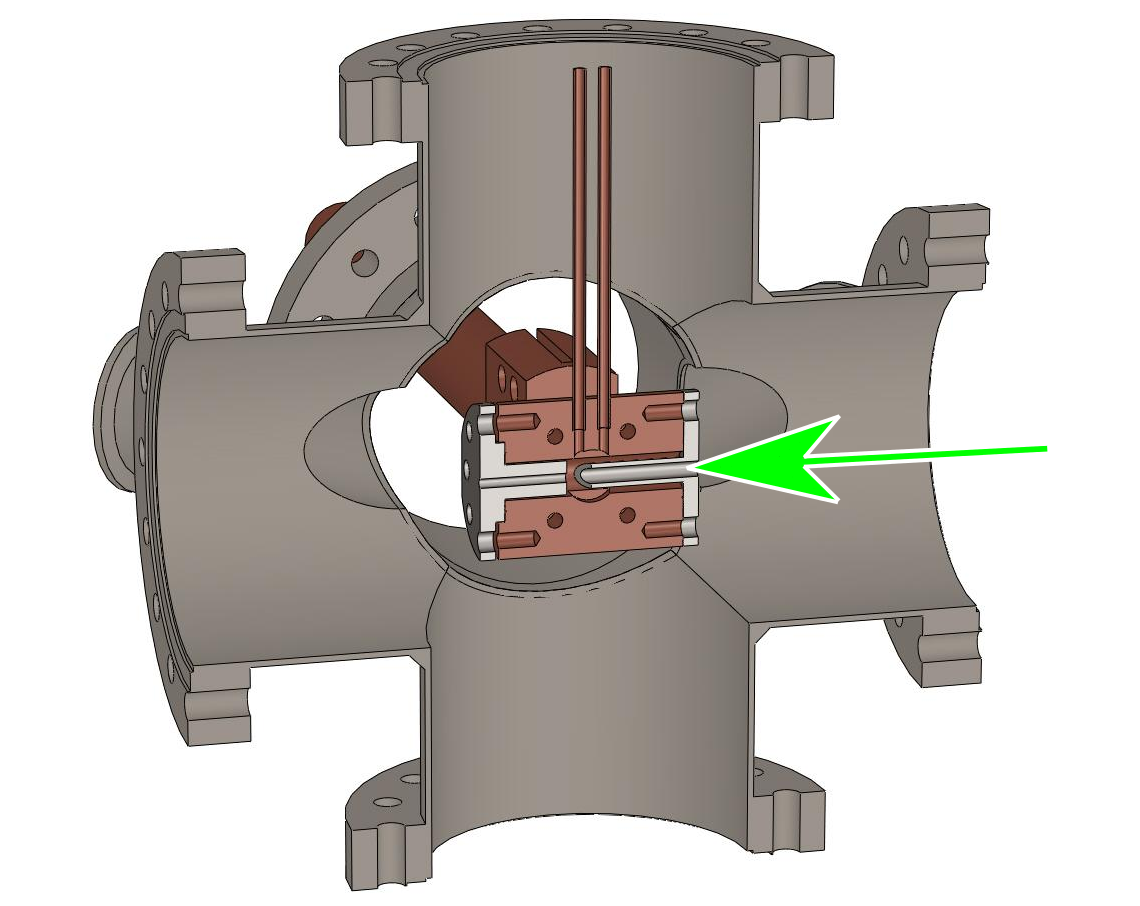}
	\caption{\label{fig::kreuzstueck} \textit{Cross-sectional view of the target cell in the surrounding CF-100 cross piece. In the experiments the ion beam is coming from the right (indicated by the arrow; green in color online), the scintillation light leaves the cell towards the observer. The connection to the dewar holding the liquid nitrogen supply can be seen on the backside. The two tubes soldered into the cell from top serve as gas inlet and outlet, respectively.}}
\end{figure}

In front of the cell (not shown in fig. \ref{fig::aufbauzelle}) an electrically monitored tantalum aperture with a borehole of 3\,mm diameter was placed. This aperture cut off the halo of the ion beam; the monitoring was used for optimal beam adjustment. Except for the titanium foil which was glued onto the beam entrance flange with an epoxy glue all parts of the cell were indium sealed. This ensured highest purity of the gas and avoided cold leaks. On the side opposite to the light detector the cell was connected to a reservoir of LN2 by a copper rod to cool it down below the condensation temperature of LAr, see fig. \ref{fig::kreuzstueck}. A 50\,$\Omega$-resistor which was soldered onto that rod for heating allowed a fine adjustment of the cell's temperature.

In addition to the connection to the dewar and the MgF$_2$-window the cell had two boreholes. Here, two copper tubes were soldered in, which served as gas inlet and outlet, respectively. These tubes both had a length of 3\,m and were in thermal contact to each other, thereby acting as heat exchanger. The temperature of both tubes as well as the temperature of the cell and the tantalum aperture were monitored by Pt100 resistance thermometers.

During operation LAr was continuously evapored and condensed, and circulated by a metal-bellows pump in an external all metal-sealed gas system. For a schematic drawing see ref. \cite{neumeier}. This gas system also contained a rare gas purifier (SAES Getters, MonoTorr$^{\text{\textregistered}}$ Phase II, PS4-MT3) which was capable of removing all chemical impurities except other rare gases. Typically, the Ar gas used was cleaned for several hours before starting the condensation process to achieve the highest purity possible. In addition, a 10\,l expansion volume was installed in the gas system to store purified argon gas. Hence, no gas from the bottle had to be refilled during the measurements and the gas was kept at a high purity. Argon from Linde AG with a purity of 99.998\% prior to purification was used.

\begin{figure}
	\centering
	\includegraphics[width=\columnwidth]{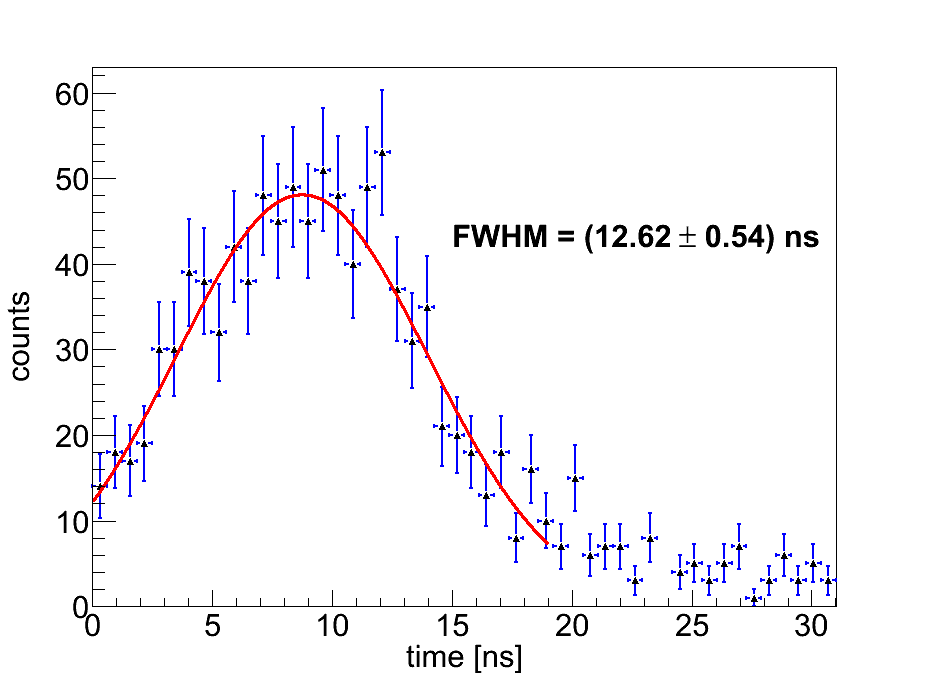}
	\caption{\label{fig::beampulse} \textit{Beam pulse profile of the pulsed sulfur beam as provided by the Munich Tandem Accelerator MLL. The time structure of the pulse was measured via X-ray emission from a Faraday cup moved into the beam line. The data can well be fitted with a Gaussian (solid line; red in color online). The fit yields a FWHM of the beam pulse of $\sim$\,12.6\,ns, taking the calibration of the TAC scale into account.}}
\end{figure}

At the beam line setup the target cell was mounted inside an evacuated CF-100 cross piece (see fig. \ref{fig::kreuzstueck}), which was connected directly to the beam line of the Munich Tandem Accelerator (MLL). The ion source of this accelerator is capable of injecting almost all kinds of ions; the accelerator itself was operated at a maximal voltage of 12\,MV. It can provide both a continuous and a pulsed ion beam, the latter with pulse widths between 5 and 13\,ns; see fig. \ref{fig::beampulse}. The continuous beam is hereby cut by a beam chopper and the single beam pulses are further compressed by a buncher system. The beam chopper delivers a logic signal each time it lets a beam pulse pass, which can be used for triggering the read-out system.

For wavelength-resolved studies in the VUV and UV range the scintillation light from the target cell was collected by an evacuated mirror optics and focussed onto the entrance slit of a VUV-reflection grating monochromator (McPherson model 218) with a focal length of f\,=\,30\,cm. The mirror optics was designed to match the aperture ratio of the monochromator of $\frac{\text{D}}{\text{f}}$\,=\,$\frac{1}{5}$, where D is the aperture diameter. The monochromator has a criss-cross Czerny-Turner design; the grating used had 1200 lines per millimeter and a blaze wavelength of 150\,nm. The optical resolution was mainly determined by the width of the entrance slits. The latter was typically chosen as 100\,\textmu m, which translated into a resolution of $\Delta\lambda$\,=\,0.3\,nm. The diffracted light was detected with a VUV-sensitive photomultiplier tube (PMT; EMI 9426 with MgF$_2$-entrance window and S20-multi-alcali cathode). 
A filter wheel between mirror optics and monochromator allowed to cut out higher diffraction orders of the grating. An aluminium plate in the filter wheel allowed to measure the dark count rate.

The same setup was used for time-resolved measurements. The monochromator was set to a fixed but selectable wavelength and the time difference between a trigger signal provided by the beam chopper of the Tandem Accelerator and the arrival time of the first scintillation light at the PMT was measured using a time-to-amplitude coverter (TAC). The start signal for this time measurement came from the PMT, the stop signal was the signal from the beam chopper which was delayed with a precision gate generator. Thus, the time resolved spectrum was recorded invertedly, however, it was ensured that each start signal has a valid stop signal. 
The intensity of the scintillation light was hereby lowered so much that only about every 100$^{\text{th}}$ pulse caused a signal from the PMT. In this way, also the delayed components of the scintillation light could be measured. The TAC-ranges used here were 500\,ns and 5\,\textmu s. 

With the aluminium plate turned in at the position of the filter wheel in order to block the direct path for light and a Faraday cup which was moved into the beam line the shape and width of the beam pulse could be measured. With this configuration only X-ray emission from the cup being hit by the ion beam was able to induce PMT signals by directly travelling through the beam line walls and hitting the photocathode. Recording a time resolved spectrum thus allowed to directly measure the shape of the beam pulse, as X-ray emission happens rather fast ($\lesssim$\,ps \cite{xray}). An example of such a measurement is shown in figure \ref{fig::beampulse}.

Both for time- and wavelength-resolved studies in the VUV and UV three different ion beams were used: protons with an energy of 10\,MeV, a 120\,MeV sulfur beam with a charge state in the vacuum\footnote{The vacuum charge state is not so important, as the ions have a velocity-dependent effective charge state in liquid argon that is independent of the vacuum charge state.} of 10+, and gold ions (Au$^{14+}$) with 195\,MeV. For all of these ion beams electronic stopping still dominates, see discussion in section \ref{sec::results}. These three ion species have been chosen to study the dependence of the scintillation light emission of liquid argon on the mass number of the incident particle. Protons are the lightest, gold ions nearly the heaviest particles available at the MLL accelerator. 
In principle, it would have been desirable to study an argon-ion beam, but since argon is a rare gas such a beam is very difficult to prepare at a tandem accelerator. Therefore, sulfur ions have been used instead, as they have nearly the same mass as argon ions, but are much easier to prepare.

For the studies of the scintillation light at longer wavelengths (250 - 970\,nm) the mirror optics and the monochromator with the PMT were replaced by a UV-Vis grating spectrometer (OceanOptics QE65000). In this setup the light was collected by a lens in a vacuum-tight flange and fed into the spectrometer with a quartz fiber. Higher diffraction orders of the light were cut out by internally installed filters. The UV-Vis spectrometer had a resolution of $\Delta\lambda$\,=\,1.4\,nm (FWHM) at 700\,nm. Time-resolved measurements were not possible with this setup, as the UV-Vis spectrometer records the full spectrum at once but without time information for the single photon hits. 
For studying the long-wavelength part of the scintillation spectrum again protons with an energy of 10\,MeV and sulfur ions (S$^{9+}$) with 115\,MeV were used. During the measurements with the VUV-sensitive setup the gold beam deposited much energy in the titanium foil and caused it to melt. Therefore, an oxygen-ion beam (O$^{6+}$ with 70\,MeV) has been used in the UV-Vis measurements instead of gold. As the UV-Vis spectrometer did not allow time-resolved measurements, these three ion beams were only used in continuous mode.

\section{Results and Discussion}
\label{sec::results}

\subsection{Scintillation Mechanism of Liquid Argon}
\label{sec::mechanismus}

The spectral features of the scintillation light spectrum emitted by LAr have been found to be similar to that of the gas phase (see e.g. \cite{heindl_epl,jortner,cheshnovsky}). Therefore, in the following we will use a nomenclature which is common for the gas phase to describe also the ''gas kinetic'' processes leading to light emission of LAr, being aware that this terminology might be partly inadequate for the liquid phase.

The energy deposited by a charged particle in LAr is converted into highly excited (Ar$^{**}$) and ionized states of argon atoms. For light incident particles like $\beta$-rays and protons, as well as heavy ions with kinetic energies $\frac{\text{E}}{\text{m}}$ much larger than $\sim$\,1\,$\frac{\text{MeV}}{\text{amu}}$ processes of electronic stopping dominate \cite{northcliffe_schilling}. The stopping power, and thus the linear energy transfer (LET) for electronic stopping can approximately be calculated using Bethe's stopping power formula \cite{bethe,ahlen}. For non-relativistic particles the electronic stopping power is a function of the square of the particle's velocity, and hence a function of $\frac{\text{E}}{\text{m}}$. For heavy ions and nuclear recoils below $\sim$\,100\,$\frac{\text{keV}}{\text{amu}}$ kinetic energy nuclear stopping dominates \cite{northcliffe_schilling}.

For electrons as incident particles it has been found that the probability to form multiply ionized argon atoms or excited argon ions is small \cite{chepel}. In contrast, low-energy heavy ions have much higher yields for the formation of highly ionized states \cite{cocke,kelbch,olson,kelbch_2}.

In recombination processes and collisions with other argon atoms the excitation energy finally leads to the formation of neutral argon excimer molecules \cite{doke_2}, Ar$_2^*$, i.e. molecules which are only strongly bound in their excited state. In the liquid and solid phase this process is also referred to as self-trapping \cite{doke88,reimann}. The typical time scale for the formation of excimer molecules in LAr is ps \cite{hitachi}; the thermalization of hot electrons takes several hundreds of ps \cite{sowada}, leading to recombination times in the ns range \cite{hitachi}. Direct radiative transitions of argon atoms in the lowest excited state to the ground state with photon energies corresponding to wavelengths of the resonance lines of 104.8\,nm and 106.7\,nm \cite{nist} are strongly suppressed in a dense medium.

Depending on the spin orientation of the excited electron in the progenitor state of the excimer molecule either singlet  ($^1\Sigma_u^+$) or triplet ($^3\Sigma_u^+$) molecules are formed. The decay of the low-lying vibrational states of these molecules to the repulsive ground state ($^1\Sigma_g^+$) gives rise to the so-called second excimer continuum, a structureless continuum centered around $\sim$\,127\,nm in LAr with a width of 7-10\,nm \cite{heindl,heindl_epl}. In the gas phase this width is found to decrease with decreasing temperature \cite{heindl,prem,morikawa,efthimiopoulos,cheshnovsky_temp}. The energy level of the singlet lies thereby slightly higher by $\sim$\,75\,meV \cite{morikawa} than the energy level of the triplet state, however, due to the width of the second excimer continuum the two transitions cannot be resolved spectroscopically.

The decay of excimer molecules which are not vibrationally relaxed leads to the emission of the so-called first excimer continuum at shorter wavelengths and the so-called classical left turning point (LTP) at longer wavelengths \cite{rhodes}. The gas kinetic processes following the formation of doubly or even more highly charged argon ions (for a possible interpretation in the gas phase see e.g. the work by Wieser et al. \cite{wieser} and references therein) lead to the emission of the so-called third excimer continuum, which is found at wavelengths longer than 170\,nm for argon and has very fast emission time constants (few ns). As the wavelengths of all these optical transitions of excimer molecules are detuned from the resonance lines, no radiative trapping occurs, and the scintillation light is free to leave the LAr. 
In the gas phase at 300\,mbar the third excimer continuum is found to be more intense with ion beam excitation than with electron beam excitation \cite{ulrich_laser}, an observation which provides a possibility for particle discrimination using the wavelength information of the scintillation light only.

The time constants for the singlet and triplet decay of the neutral excimer molecules in the liquid phase are very different: light emission in allowed singlet decays typically happens within a few nanoseconds, while for triplet decays life times of more than 1\,\textmu s have been reported \cite{heindl,heindl_epl,peiffer,carvalho} in the absence of impurities. The direct transition from triplet states into the ground state is forbidden, and only mixing between $^3\Sigma_u^+$ and $^1\Pi_u$ states through spin-orbit coupling \cite{kubota} renders the decay possible.
The intensity ratio of singlet and triplet light is found to be dependent on the type and energy of the incident particle \cite{heindl,heindl_epl,peiffer,lippincott,morikawa,carvalho,hitachi_83}. The time constants for the singlet and triplet light are, however, very similar for different incident particles \cite{peiffer,lippincott,carvalho,hitachi_83}, indicating that the same species are emitting the scintillation light and demonstrate the absence of substantial non-radiative quenching \cite{chepel}. The values quoted in literature are also summarized in table \ref{tab::liter}.

\begin{table*}[htbp]
	\centering
	\begin{tabular}{|>{\columncolor{hellgrau}}c|c|c|c|c|}
		\hline \rowcolor{dunkelgrau} &&&&\\ \rowcolor{dunkelgrau} &&&&\\
		\multirow{-3}{2cm}{\centering \cellcolor{dunkelgrau} Exciting\\Particle} & \multirow{-3}{2.5cm}{\centering \cellcolor{dunkelgrau} $\tau_{\text{s}}$\\$[$ns$]$} & \multirow{-3}{2.7cm}{\centering \cellcolor{dunkelgrau} $\tau_{\text{t}}$\\$[$ns$]$} & \multirow{-3}{2cm}{\centering \cellcolor{dunkelgrau} $\frac{\text{I}_{\text{s}}}{\text{I}_{\text{t}}}$} & \multirow{-3}{2cm}{\centering \cellcolor{dunkelgrau} Ref.} \\
		\hline \hline &&&&\\
		\multirow{-2}{2cm}{\centering \cellcolor{hellgrau} $\gamma$} & \multirow{-2}{2.5cm}{\centering 1.65\,$\pm$\,0.1} & \multirow{-2}{2.7cm}{\centering -} & \multirow{-2}{2cm}{\centering -} & \multirow{-2}{2cm}{\centering \cite{morikawa}}\\
		\hline &&&&\\
		\multirow{-2}{2cm}{\centering \cellcolor{hellgrau} $\gamma$} & \multirow{-2}{2.5cm}{\centering -} & \multirow{-2}{2.7cm}{\centering 1463\,$\pm$\,55} & \multirow{-2}{2cm}{\centering -} & \multirow{-2}{2cm}{\centering \cite{lippincott}}\\
		\hline &&&&\\
		\multirow{-2}{2cm}{\centering \cellcolor{hellgrau} $\gamma$} & \multirow{-2}{2.5cm}{\centering 10\,$\pm$\,5} & \multirow{-2}{2.7cm}{\centering 1200\,$\pm$\,20} & \multirow{-2}{2cm}{\centering 0.30\,$\pm$\,0.01} & \multirow{-2}{2cm}{\centering \cite{peiffer}}\\
		\hline &&&&\\
		\multirow{-2}{2cm}{\centering \cellcolor{hellgrau} e$^-$} & \multirow{-2}{2.5cm}{\centering $<$\,6.2} & \multirow{-2}{2.7cm}{\centering 1300\,$\pm$\,60} & \multirow{-2}{2cm}{\centering 0.51\,$\pm$\,0.05} & \multirow{-2}{2cm}{\centering \cite{heindl,heindl_epl}}\\
		\hline &&&&\\
		\multirow{-2}{2cm}{\centering \cellcolor{hellgrau} e$^-$} & \multirow{-2}{2.5cm}{\centering 4.6} & \multirow{-2}{2.7cm}{\centering 1540} & \multirow{-2}{2cm}{\centering 0.26} & \multirow{-2}{2cm}{\centering \cite{carvalho}}\\
		\hline &&&&\\
		\multirow{-2}{2cm}{\centering \cellcolor{hellgrau} e$^-$} & \multirow{-2}{2.5cm}{\centering 6\,$\pm$\,2} & \multirow{-2}{2.7cm}{\centering 1590} & \multirow{-2}{2cm}{\centering 0.3} & \multirow{-2}{2cm}{\centering \cite{hitachi_83}}\\
		\hline &&&&\\
		\multirow{-2}{2cm}{\centering \cellcolor{hellgrau} $\alpha$} & \multirow{-2}{2.5cm}{\centering 4.4} & \multirow{-2}{2.7cm}{\centering 1100} & \multirow{-2}{2cm}{\centering 3.3} & \multirow{-2}{2cm}{\centering \cite{carvalho}}\\
		\hline &&&&\\
		\multirow{-2}{2cm}{\centering \cellcolor{hellgrau} $\alpha$} & \multirow{-2}{2.5cm}{\centering 7.1\,$\pm$\,1.0} & \multirow{-2}{2.7cm}{\centering 1660\,$\pm$\,100} & \multirow{-2}{2cm}{\centering 1.3} & \multirow{-2}{2cm}{\centering \cite{hitachi_83}}\\
		\hline &&&&\\
		\multirow{-2}{2cm}{\centering \cellcolor{hellgrau} $\alpha$} & \multirow{-2}{2.5cm}{\centering 10\,$\pm$\,5} & \multirow{-2}{2.7cm}{\centering 1200\,$\pm$\,20} & \multirow{-2}{2cm}{\centering 2.6\,$\pm$\,0.1} & \multirow{-2}{2cm}{\centering \cite{peiffer}}\\
		\hline &&&&\\
		\multirow{-2}{2cm}{\centering \cellcolor{hellgrau} n} & \multirow{-2}{2.5cm}{\centering 10\,$\pm$\,5} & \multirow{-2}{2.7cm}{\centering 1200\,$\pm$\,20} & \multirow{-2}{2cm}{\centering 3.5\,$\pm$\,0.2}& \multirow{-2}{2cm}{\centering \cite{peiffer}}\\
		\hline &&&&\\ &&&&\\
		\multirow{-3}{2cm}{\centering \cellcolor{hellgrau} fission\\fragments} & \multirow{-3}{2.5cm}{\centering 6.8\,$\pm$\,1.0} & \multirow{-3}{2.7cm}{\centering 1550\,$\pm$\,100} & \multirow{-3}{2cm}{\centering 3} & \multirow{-3}{2cm}{\centering \cite{hitachi_83}}\\
		\hline  
	\end{tabular}
	\caption{\label{tab::liter}\textit{Literature values \cite{heindl,heindl_epl,peiffer,lippincott,morikawa,carvalho,hitachi_83} for the light-emission time constants of singlet and triplet liquid argon excimer decays as well as the light-intensity ratio $\frac{I_s}{I_t}$ quoted for different exciting particles.}}
\end{table*}

A possible explanation for these experimental observations is a redistribution of excitation energy between the two states in collisions with hot electrons in the early phase after the excitation \cite{lorents},

\begin{equation}
	^1\Sigma_u^+ + e^- \leftrightarrow ^3\Sigma_u^+ + (e^-)' , \label{eq::mixing}
\end{equation}

and the much faster decay of the singlet. As the electron density, and consequently the probability for collisions, is different for different particles \cite{chepel} the intensity ratio between singlet and triplet decays provides an applicable parameter for particle discrimination. This has been shown in studies on the pulse shape in wavelength-integrated measurements \cite{peiffer,lippincott}. However, the fast light emission from the third excimer continuum at longer wavelengths might mimic the fast singlet component in such measurements, influencing the discrimination potential. In order to tap the full potential of LAr wavelength-resolved studies are mandatory.

\subsection{Gas Preparation and Beam Settings}

Prior to all measurements the argon gas used was purified for several hours to get rid of all chemical impurities, which would severely disturb the measurements \cite{heindl_epl,heindl_jinst,rhodes}. The purity of the gas was monitored by looking for the 557.7\,nm-emission line caused by oxygen \cite{rhodes,oxyline}. Only after this line had completely vanished the cool-down process was started. As mentioned above the rare gas purifier was not able to remove other rare gases from argon, xenon in particular. As a result, all spectra recorded within the scope of this work show a very weak emission line stemming from xenon at 148.9\,nm \cite{heindl_epl,heindl_jinst}. 
From a comparison of the emission spectra in the gas phase to argon spectra where quantitatively defined traces of xenon were deliberately added \cite{efthimiopoulos} the remaining xenon concentration in our gas could be estimated to be much smaller than 3\,ppm; a comparison in the liquid phase to spectra recorded with electron beam excitation \cite{heindl,heindl_epl,heindl_jinst} yielded a xenon impurity of even less than 0.3\,ppm.

The beam current of the MLL ion beam sometimes showed some instabilities. In order to prevent the wave\-length-resolved spectra to be spoiled by such fluctuations the beam current was monitored for several minutes before and after recording each spectrum. Only in case that no fluctuations beyond a 10\%-level were found, the spectrum was used in the final analysis. The beam currents were in all cases set to values which were low enough not to make the LAr boil. This was monitored by the 696.5\,nm emission line \cite{heindl_jinst,nist} (4p-4s transition line of ArI), which appears only for gaseous argon and is strongly suppressed in the liquid phase. These beam current settings also avoided multiple excitations of one argon atom: 
for typical values of the beam current ($\sim$\,5\,nA for the proton beam, $\sim$\,6\,nA for the sulfur ions, and $\sim$\,0.3\,nA for the gold ions) the calculated densities of excited and ionized states were about 8 orders of magnitude lower than the density of LAr. All spectra presented in the following stem consequently from argon atoms which have interacted with one single beam ion only.

\subsection{Wavelength-Resolved Studies}
\label{sec::wavelengthstudies}

The wavelength-resolved scintillation-light spectra of argon have been re\-cor\-ded with the McPherson monochromator in a wavelength interval from 110\,nm through 310\,nm with the three different ion beams, both for the liquid phase (T\,=\,(86.1\,$\pm$\,1.2)\,K; p\,=\,970\,mbar) and cold gas (T\,= (98.8\,$\pm$\,1.2)\,K; p\,=\,1200\,mbar). The latter are of particular interest for two-phase detectors, which use both the scintillation light of argon in the liquid and the gas phase \cite{ArDM,WARP,darkside}. For the wavelength-resolved studies continuous ion beams from the accelerator have been used.

The raw data had to be corrected for a diminishing transmission of the MgF$_2$-exit window due to condensation of residual gas atoms on its surface. This correction was carried out analogously to the procedure described in \cite{neumeier}. However, as the correction function was only known down to 118\,nm all spectra shown in the following end at this wavelength.

In addition, the detector response function was fully taken into account. This function was derived from a comparison of a spectrum recorded in the gas phase at 300 mbar at the beam line setup and an absolutely calibrated spectrum recorded at the same pressure \cite{hofmann}. Corrections for light reflections at the surfaces of the MgF$_2$-window (Fresnel formula) were not taken into account, as the indices of refraction of both MgF$_2$ and LAr are not precisely determined in the VUV wavelength region. However, the influence of these corrections on the spectral shape is expected to be small \cite{hofmann}. Furthermore, light attenuation in the window was also not taken into account. The attenuating effects by filters used at longer wavelengths to cut out higher diffraction orders of the scintillation light at short wavelengths were corrected by scaling the spectra appropriately to match in the overlap wavelength region. 
The detector's dark count rate, which was determined by blocking the light by the aluminium plate in the filter wheel, was subtracted from the spectra. In the measurements with the monochromator and the PMT this dark count rate was about 270\,counts per bin, which gives a minimum signal-to-noise ratio of about one for the weakest light emission at $\sim$\,240\,nm. For the second excimer continuum a signal-to-noise ratio of 400 was exceeded.

\begin{figure}
	\centering
	\includegraphics[width=\columnwidth]{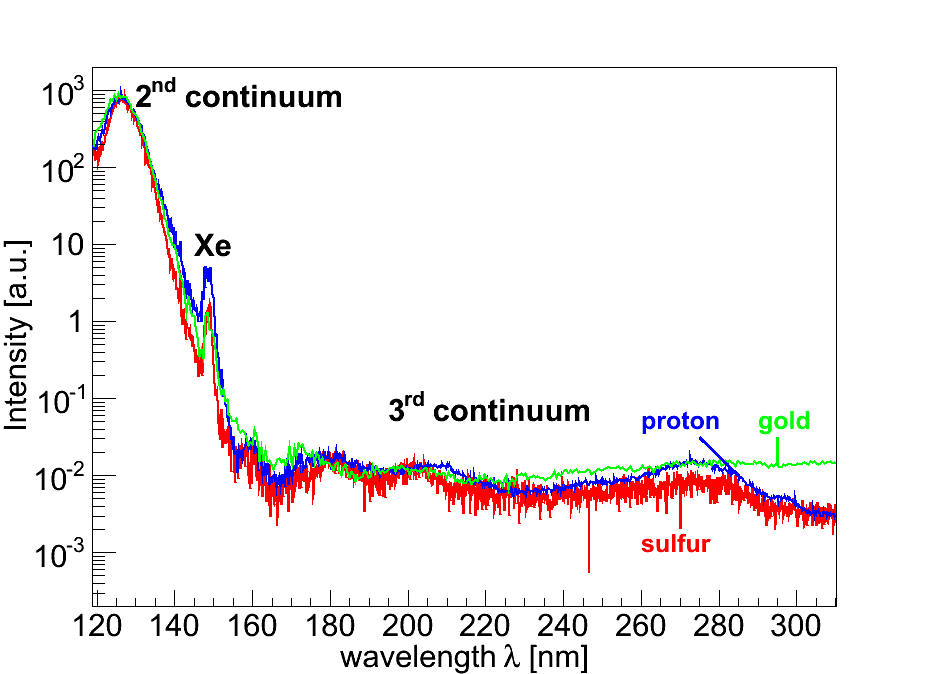}
	\caption{\label{fig::wellenlaengenspektren} \textit{Wavelength-resolved scintillation light spectra of LAr in the wavelength range from 118\,nm through 310\,nm when excited with different ion beams (sulfur, red in color online, protons in blue, and gold ions in green). The single spectra were corrected for dark count rate, detector response function and the transmission of the MgF$_2$-window, and scaled to match each other at the peak emission wavelength of the second excimer continuum at $\sim$\,127\,nm. The resolution is $\Delta\lambda$\,=\,0.3\,nm. For details see text.}}
\end{figure}

Figure \ref{fig::wellenlaengenspektren} shows the wavelength-resolved scintillation light spectra of LAr when excited with different ion beams. The by far dominant emission feature is the second excimer continuum at wavelengths shorter than 145\,nm. It peaks at (126.8\,$\pm$\,0.1)\,nm for the proton beam, (126.4\,$\pm$ 0.1)\,nm for the sulfur beam, and (125.8\,$\pm$\,0.2)\,nm for the gold beam. Towards longer wavelengths, at 148.9\,nm, the impurity emission of xenon appears, the only emission line visible in the spectra. The structure around 160\,nm is likely attributed to the analogue of the classical LTP in the liquid phase. For wavelengths longer than $\sim$\,170\,nm various emission features of the third excimer continuum follow. In dedicated measurements with highest resolution possible all these emission features are found to be structureless continua. 
For the excitation with gold ions a flat continuum likely due to bremsstrahlung by secondary electrons can be seen above $\sim$\,220\,nm.

The spectra obtained with the different incident particles are very similar. In particular, they show the same spectral emission features and the same intensity ratio between second and third excimer continuum. The differences which could be expected from the studies in the dilute gas phase \cite{ulrich_laser} in the wavelength region of the third excimer continuum do not show up in the liquid phase. This could have two reasons: On one hand, the number of progenitor states of the third excimer continuum compared to the number of excimer states leading to the emission of the second excimer continuum might be equal for the three different exciting particles, leading to equal intensities. On the other hand, the wavelength-resolved scintillation-light spectra could also be dominated by light emission from secondary electrons released by the ion beams, leading to a similar spectral shape. The similarity of the spectra makes particle discrimination in LAr impossible using the wavelength-resolved information 
alone.

\begin{figure}
	\centering
	\includegraphics[width=\columnwidth]{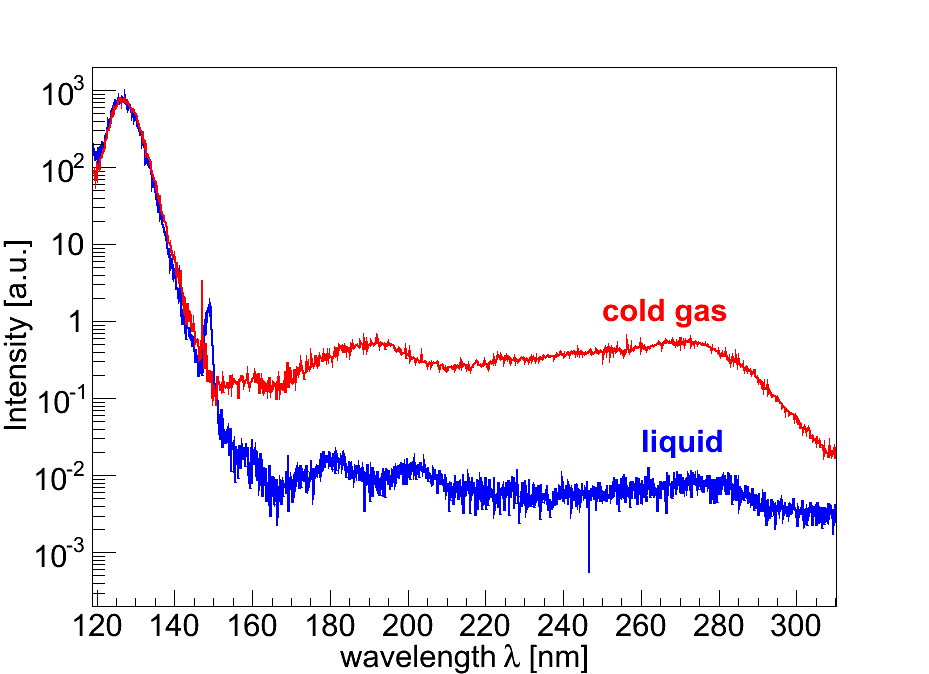}
	\caption{\label{fig::gasfluessig} \textit{Wavelength-resolved scintillation light spectra of argon in liquid (blue in color online) and gas phase (1200\,mbar, $\sim$\,99\,K; red in color online) in a wavelength range from 118\,nm to 310\,nm when excited by sulfur ions. The two spectra were corrected for dark count rate, detector response function and the transmission of the MgF$_2$-window, and scaled to match each other at the peak emission wavelength of the second excimer continuum at $\lambda\,\sim$\,127\,nm. The resolution is $\Delta\lambda$\,=\,0.3\,nm. For details see text.}}
\end{figure}

Compared to the gas phase (see fig. \ref{fig::gasfluessig}) the intensity in the wavelength region of the third excimer continuum is strongly suppressed for LAr: in the liquid phase only 0.0094\% of the light (0.016\% of the photons) is emitted between 170\,nm and 310\,nm. In the gas phase at $\sim$\,99\,K, 0.43\% of the light intensity (0.77\% of the scintillation photons) are emitted in the given wavelength region. This could be explained if the progenitor states of the third excimer continuum, highly charged argon ions \cite{wieser,langhoff,treschalov}, do not exist in the liquid phase long enough to form ionic excimer molecules, and may alternatively undergo, e.g., charge transfer reactions with neighboring atoms. Compared to this, 99.9\% of the scintillation light of LAr (and also 99.9\% of the photons) are emitted in the wavelength region of the second excimer continuum.

The peak wavelengths and shapes of the second excimer continua are found to be independent of the state of aggregation. The xenon impurity emission is found at 146.96\,nm (XeI line \cite{nist}) in the gas phase and has a width which is smaller than the instrumental resolution. As no\-ted above, in LAr this line is shifted to 148.9\,nm, an effect also reported in the literature \cite{cheshnovsky}, and it is considerably broadened to (1.8\,$\pm$\,0.4)\,nm FWHM.

As for LAr also for argon in the gas phase the scintillation light spectra obtained for different incident ion beams turn out to be similar. In particular, the same emission features are found, and the ratio between second and third excimer continuum is equal. Once again, this does not permit particle discrimination using the wavelength-resolved light information only. Even in the gas phase the time-integrated amount of scintillation light emitted in the wavelength region of the third continuum is tiny. In our studies we find the width of the second excimer continuum to be slightly different for the different ion beams (see fig. \ref{fig::wellenlaengenspektren}), which could be related to a different local heating. In the gas phase the width is known to be influenced by temperature \cite{heindl,prem,morikawa,efthimiopoulos,cheshnovsky_temp}.

The scintillation spectra of gaseous and liquid argon at longer wavelengths (250\,nm through 970\,nm) were recor\-ded with the UV-Vis grating spectrometer. Also in this experiment the raw data were corrected for the detector response \cite{hofmann}, however, no corrections for reflections at the window's surfaces or a diminishing transmission were made. The latter were found to play no important role for wavelengths longer than 250\,nm \cite{neumeier}. Background due to stray light, X-rays and dark noise was subtracted; in each case it was thoroughly checked that the parameter settings of the spectrometer and the ion beam did not influence the shape of the spectra.

\begin{figure}
	\centering
	\includegraphics[width=\columnwidth]{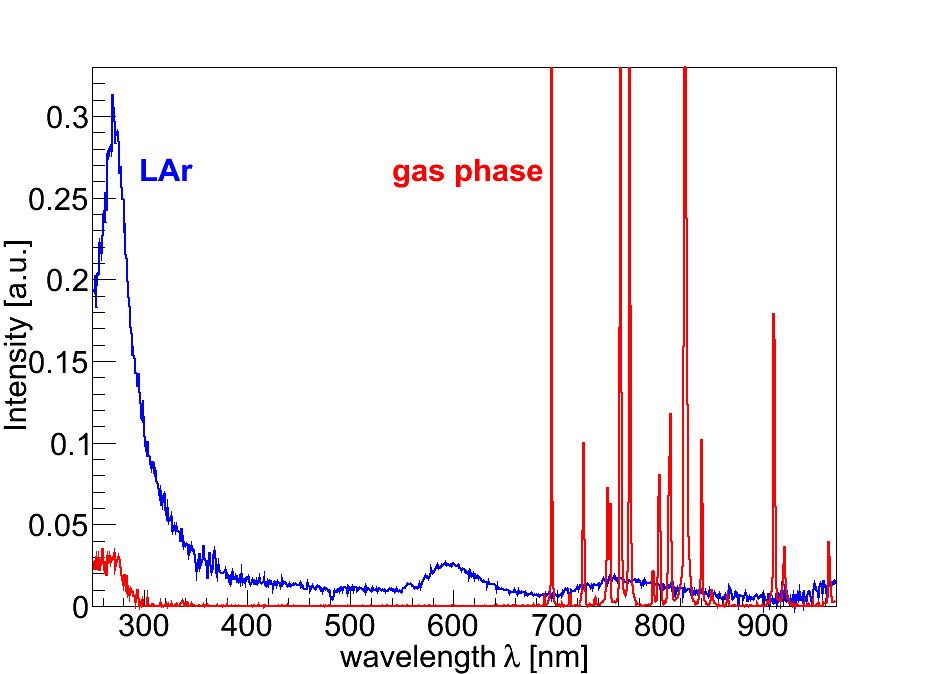}
	\caption{\label{fig::oo_spectren} \textit{Wavelength-resolved scintillation light spectra of gaseous and liquid argon recorded with the UV-Vis spectrometer and proton beam excitation. The spectrum of the gas phase (red in color online) was recorded at $\sim$\,90\,K and 1200\,mbar with a beam current of 7\,nA, the spectrum of LAr (blue in color online) with a beam current of 5\,nA but a six times longer integration time. The spectra are not scaled. The rise in intensity for wavelengths longer than $\sim$\,940\,nm in the spectrum recorded for LAr is caused by applying the detector response function. For details see text.}}
\end{figure}

Figure \ref{fig::oo_spectren} shows the wavelength-resolved spectra of gas\-eous and liquid argon from the UV to the near IR range. For the gas phase most of the light is emitted in the line radiation of ArI. Below 300\,nm, only a low-intensity emission continuum is found which can possibly be attributed to transitions of ArIV \cite{heindl} or is the long-wavelength part of the third excimer continuum as seen in fig. \ref{fig::gasfluessig}. An analogue continuum appears in LAr. However, the line radiation has vanished, but a broad continuum around 755\,nm is found. Besides this, another continuum centered at 595\,nm can be seen, which is X-ray-induced fluorescence of MgF$_2$ (unpublished data). The tiny dip at 480\,nm is an artifact of the stray light correction; the tiny structure around 558\,nm is caused by a residual oxygen impurity \cite{oxyline}. 
The underlying broad continuum emission can be attributed to bremsstrahlung from secondary electrons, and is enhanced with increasing beam power. The rising intensity starting from $\sim$\,940\,nm is most likely an artifact caused by the detector response function. Results on the light emission of LAr in the infrared region will be published in a forthcoming paper.

The total light intensity of LAr for wavelengths longer than $\sim$\,300\,nm is very low and, furthermore, again very similar for different incident ions. Particle discrimination is therefore not possible using the spectral information of the scintillation light only.

\subsection{Time-Resolved Studies}
\label{sec::timestudies}

\begin{figure}
	\centering
	\includegraphics[width=\columnwidth]{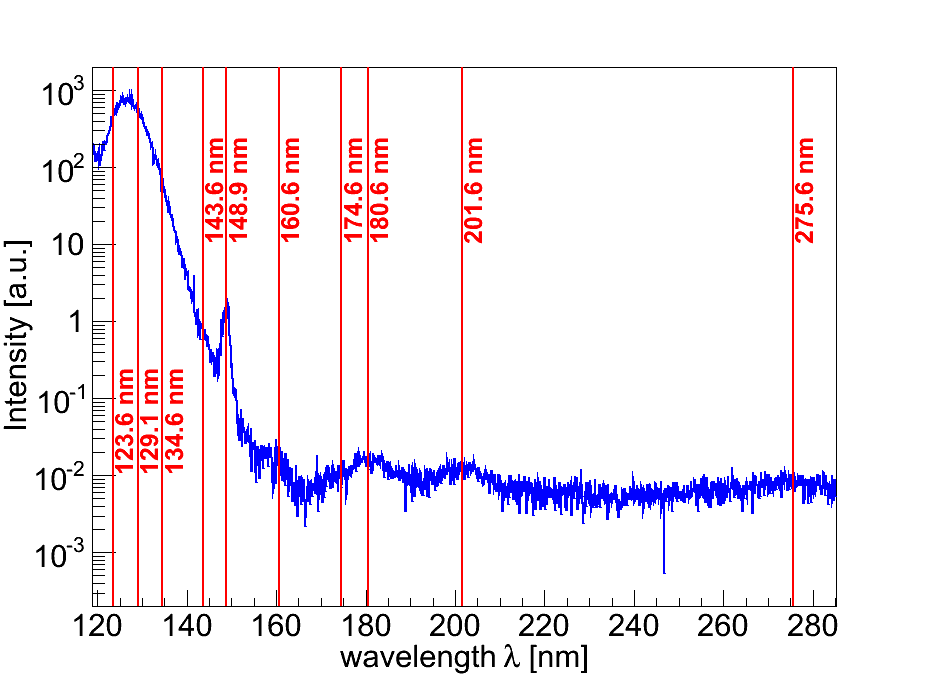}
	\caption{\label{fig::wellenlaengenuebersicht} \textit{Scintillation light spectrum of LAr recorded with sulfur beam excitation (blue in color online). The solid vertical lines (red in color online) indicate the wavelengths which have been investigated in time-resolved measurements; the figures give the exact wavelengths. All features of interest of the spectrum were covered, in particular the second excimer continuum including its blue and red wings. The wavelength-resolved spectrum was only recorded down to 118\,nm as the correction for the diminishing transmission was only known down to this wavelength, however, at 115.6\,nm a further time-resolved spectrum was recorded.}}
\end{figure}

The time structure of the scintillation light emitted from LAr was investigated for a set of dedicated wavelengths (see fig. \ref{fig::wellenlaengenuebersicht}) using the method described in section \ref{sec::setup}. For these data, corrections for detector response or transmission of the exit window were unneccessary, as only the relative time structure at one certain wavelength was investigated. The only correction which had to be taken into account was the time range calibration of the TAC.

At first, the shape of the beam pulses had to be measured. This was carried out using the method described in sec. \ref{sec::setup}. The beam pulses turned out to be nearly perfectly Gaussian shaped (see fig. \ref{fig::beampulse}). For the sulfur beam a FWHM of this beam pulse of (12.62\,$\pm$\,0.54)\,ns was found, for the other two projectiles similar but slightly smaller widths were obtained: (7.14\,$\pm$\,0.12)\,ns for the gold beam and (5.46\,$\pm$\,0.05)\,ns for the proton beam. The precise determination of these widths was important to measure the decay time constant of the singlet decay, which is of the same order of magnitude as the width of the exciting beam pulse. 
Therefore, the first tens of nanoseconds of each time spectrum recorded were fitted by a convolution of a Gaussian and an exponential function using the toolkit program RooFit \cite{roofit}. The Gaussian accounts for the shape of the exciting beam pulse, the exponential function for the time structure of the fast component of scintillation light emission. An additive constant was used to account for the dark count rate. This fit had 5 free parameters: the center value and the normalization of the Gaussian, normalization and decay time constant of the exponential function, and the additive constant. The widths of the Gaussian-shaped function were taken from the Gaussian fits to the beam pulse profile and kept fixed.

\begin{figure}
	\centering
	\includegraphics[width=\columnwidth]{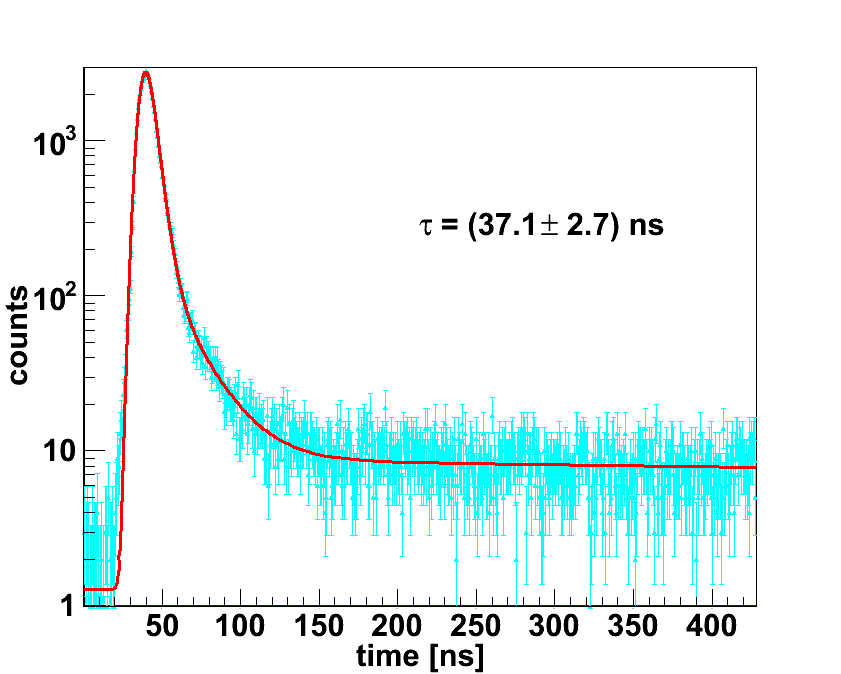}
	\caption{\label{fig::gaussexpoconvolution} \textit{Logarithmically scaled time spectrum of LAr at a wavelength of 160.6\,nm when excited with a sulfur beam. The wavelength resolution is $\Delta\lambda$\,=\,1.5\,nm. The spectrum was fitted with a convolution of a Gaussian representing the beam pulse and an exponential function with a decay time constant $\tau\,=\,(37.1\,\pm\,2.7)$\,ns (solid line; red in color online). An additive constant accounts for the dark count rate ($\approx$\,11\,ns$^{-1}$). The width of the Gaussian was kept fixed at the value derived from beam pulse measurement ((12.62\,$\pm$\,0.54)\,ns).}}
\end{figure}

In fig. \ref{fig::gaussexpoconvolution} an example of this procedure is shown. This figure depicts the time-resolved spectrum recorded at 160.6 nm, which has only one single decay time constant. The data are fitted with a convolution of a Gaussian and an exponential function plus a constant. This fit describes the data excellently.

For the time dependence of the light emission at the peak emission wavelength of the second excimer continuum as well as at its blue and red wings the following function was fitted to the tail of the time distribution of the scintillation light (t\,=\,0 is the time of highest intensity of the exciting beam pulse):

\begin{eqnarray}
	f(t) &=& \underset{singlet}{\underbrace{A_s \cdot exp\left\{-\frac{t}{\tau_s}\right\}}} + \underset{triplet}{\underbrace{A_t \cdot exp\left\{-\frac{t}{\tau_t}\right\}}} + \notag \\  
	&+& \underset{recombination}{\underbrace{\frac{A_{rec}}{\left(1+\frac{t}{\tau_{rec}}\right)^2}}} + C_0 . \label{eq::recofit}
\end{eqnarray}

The first exponential function with time constant $\tau_{\text{s}}$ accounts for the fast decay of vibrationally relaxed neutral excimer molecules in the singlet state, the second one with time constant $\tau_{\text{t}}$ for the slow decay of molecules in the triplet state. The initial value for the fit of the fast decay time constant $\tau_{\text{s}}$ was set to the outcome of fitting the first (fast) part of the time spectrum with the convolution of a Gaussian and an exponential function as described above. Typically, the value for $\tau_{\text{s}}$ as derived from fitting eq. (\ref{eq::recofit}) to the data did not change much compared to its initial value. The parameters A$_{\text{s,t}}$ scale the exponential functions; C$_0$ is an additive constant to account for the dark count rate.

\begin{figure}
	\centering
	\includegraphics[width=\columnwidth]{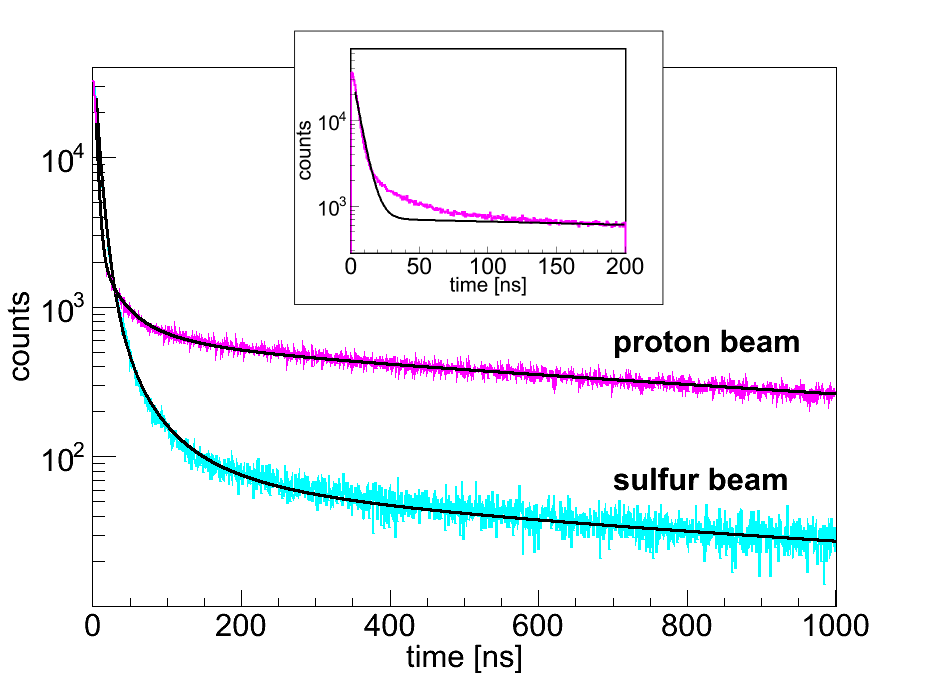}
	\caption{\label{fig::psd_p_S} \textit{Logarithmically scaled time spectrum of LAr at the peak emission wavelength of the second excimer continuum when excited by a proton beam (128.6\,nm; upper line; purple in color online) and a sulfur beam (129.1\,nm; lower line; light blue in color online). The wavelength resolution is $\Delta\lambda$\,=\,1.5\,nm. The spectra were scaled to match each other at t\,=\,0. For excitation with the proton beam the intensity of the slow component is clearly enhanced, while the fast component has a shorter time constant. A fit to each of the data is shown, too, which was performed using eq. (\ref{eq::recofit}). The inset shows the proton data, but fitted with eq. (\ref{eq::recofit}) without the recombination term. Clearly, a sum of two exponential functions and a constant does not fully describe the data at intermediate times.}}
\end{figure}

The sum of two exponential functions, however, did not perfectly describe the data at intermediate times ($\sim$\,20 - 100\,ns), see inset in fig. \ref{fig::psd_p_S}. Therefore, equation (\ref{eq::recofit}) also takes into account light emission from neutral excimer molecules formed in recombination processes of ions (or ionic excimer molecules) with electrons. For a neutral me\-dium with the same densities of electrons and ions with a homogeneous spatial charge distribution, light from these recombinations is expected to follow the third term of eq. (\ref{eq::recofit}) \cite{ribitzki}. The fit also assumed that $\tau_{\text{rec}}$ is independent of time, i.e. the electrons have already fully thermalized at the time of recombination. For LAr, thermalization times of less than 1\,ns have been reported \cite{sowada}. 
In principle, a recombination process can lead to either a molecule in a singlet or a triplet state. 
While the fast decay of singlet states gives rise to light with an emission time constant which is dominated by the time constant of the recombination, light emission from decays of triplet states mainly contributes to the total intensity of the delayed component of the time spectrum. Therefore, one time constant was sufficient to describe recombination. Two typical fits to time spectra of LAr when excited by a proton and a sulfur beam are shown in fig. \ref{fig::psd_p_S} for the second excimer continuum.

Table \ref{tab::ergebnisse} summarizes the results for the time constants $\tau_{\text{s,t,rec}}$ at the different wavelengths investigated within the second excimer continuum. The figures were prepared using the data presented in \cite{hofmann}, but re-analysed using a slightly improved fitting algorithm. It was hereby thoroughly checked that the time constants do neither depend on the width of the monochromator's entrance slit nor on the beam intensity, i.e. the light intensity. However, especially the fast decay time scale changes greatly when omitting the recombination term in eq. (\ref{eq::recofit}): for example, for the proton beam data at 128.6\,nm it changes from (3.20\,$\pm$\,0.02)\,ns when including the recombination term in the fit function to (5.07\,$\pm$0.03)\,ns when omitting it. The slow decay time scale becomes 18\% shorter; the reduced $\chi^2$ worsens from 9498/7245 to 32070/7247. As mentioned above, during the beam time with the gold beam some technical difficulties appeared concerning the 
titanium foil, which made it impossible to determine the slow time constants in that particular case. 

\begin{table}[htbp]
	\centering
	\begin{tabular}{|>{\columncolor{hellgrau}}c|c|c|c|c|}
		\hline \rowcolor{dunkelgrau} &&&&\\ \rowcolor{dunkelgrau} &&&&\\
		\multirow{-3}{1.8cm}{\centering \cellcolor{dunkelgrau} wavelength} & \multirow{-3}{1.2cm}{\centering \cellcolor{dunkelgrau} $\tau_{\text{s}}$\\$[$ns$]$} & \multirow{-3}{1.2cm}{\centering \cellcolor{dunkelgrau} $\tau_{\text{t}}$\\$[$ns$]$} & \multirow{-3}{1.2cm}{\centering \cellcolor{dunkelgrau} $\tau_{\text{rec}}$\\$[$ns$]$} & \multirow{-3}{1.2cm}{\centering \cellcolor{dunkelgrau} $\frac{\text{I}_{\text{s}}}{\text{I}_{\text{t}}}$} \\
		\hline \hline
		\multicolumn{5}{|c|}{\cellcolor{dunkelgrau} Sulfur beam} \\
		\hline \hline &&&&\\ &&&&\\
		\multirow{-3}{1.8cm}{\centering \cellcolor{hellgrau} 115.6\,nm} & \multirow{-3}{1.2cm}{\centering 5.79\\$\pm$ 0.20} & \multirow{-3}{1.2cm}{\centering 1047.5\\$\pm$ 96.7} & \multirow{-3}{1.2cm}{\centering 36.4\\$\pm$ 0.4} & \multirow{-3}{1.2cm}{\centering 3.03\\$\pm$ 0.35} \\
		\hline &&&&\\ &&&&\\
		\multirow{-3}{1.8cm}{\centering \cellcolor{hellgrau} 123.6\,nm} & \multirow{-3}{1.2cm}{\centering 7.41\\$\pm$ 0.11} & \multirow{-3}{1.2cm}{\centering 1344.9\\$\pm$ 32.0} & \multirow{-3}{1.2cm}{\centering 38.1\\$\pm$ 0.1} & \multirow{-3}{1.2cm}{\centering 3.16\\$\pm$ 0.09} \\
		\hline &&&&\\ &&&&\\
		\multirow{-3}{1.8cm}{\centering \cellcolor{hellgrau} 129.1\,nm} & \multirow{-3}{1.2cm}{\centering 6.47\\$\pm$ 0.09} & \multirow{-3}{1.2cm}{\centering 1224.0\\$\pm$ 17.9} & \multirow{-3}{1.2cm}{\centering 37.4\\$\pm$ 0.2} & \multirow{-3}{1.2cm}{\centering 2.19\\$\pm$ 0.07} \\
		\hline &&&&\\ &&&&\\
		\multirow{-3}{1.8cm}{\centering \cellcolor{hellgrau} 134.6\,nm} & \multirow{-3}{1.2cm}{\centering 5.97\\$\pm$ 0.06} & \multirow{-3}{1.2cm}{\centering 1205.3\\$\pm$ 14.8} & \multirow{-3}{1.2cm}{\centering 38.1\\$\pm$ 0.1} & \multirow{-3}{1.2cm}{\centering 1.56\\$\pm$ 0.05} \\
		\hline &&&&\\ &&&&\\
		\multirow{-3}{1.8cm}{\centering \cellcolor{hellgrau} 143.6\,nm} & \multirow{-3}{1.2cm}{\centering 5.24\\$\pm$ 0.15} & \multirow{-3}{1.2cm}{\centering 1075.2\\$\pm$ 62.1} & \multirow{-3}{1.2cm}{\centering 32.6\\$\pm$ 0.7} & \multirow{-3}{1.2cm}{\centering 2.00\\$\pm$ 0.19} \\
		\hline \hline
		\multicolumn{5}{|c|}{\cellcolor{dunkelgrau} Proton beam} \\
		\hline \hline &&&&\\ &&&&\\
		\multirow{-3}{1.8cm}{\centering \cellcolor{hellgrau} 121.6\,nm} & \multirow{-3}{1.2cm}{\centering 3.40\\$\pm$ 0.14} & \multirow{-3}{1.2cm}{\centering 1269.0\\$\pm$ 92.6} & \multirow{-3}{1.2cm}{\centering 8.2\\$\pm$ 0.1} & \multirow{-3}{1.2cm}{\centering 0.43\\$\pm$ 0.03} \\
		\hline &&&&\\ &&&&\\
		\multirow{-3}{1.8cm}{\centering \cellcolor{hellgrau} 123.6\,nm} & \multirow{-3}{1.2cm}{\centering 2.90\\$\pm$ 0.05} & \multirow{-3}{1.2cm}{\centering 1285.6\\$\pm$ 10.0} & \multirow{-3}{1.2cm}{\centering 8.0\\$\pm$ 0.1} & \multirow{-3}{1.2cm}{\centering 0.30\\$\pm$ 0.01} \\
		\hline &&&&\\ &&&&\\
		\multirow{-3}{1.8cm}{\centering \cellcolor{hellgrau} 128.6\,nm} & \multirow{-3}{1.2cm}{\centering 3.20\\$\pm$ 0.02} & \multirow{-3}{1.2cm}{\centering 1355.8\\$\pm$ 5.8} & \multirow{-3}{1.2cm}{\centering 7.7\\$\pm$ 0.1} & \multirow{-3}{1.2cm}{\centering 0.28\\$\pm$ 0.01} \\
		\hline &&&&\\ &&&&\\
		\multirow{-3}{1.8cm}{\centering \cellcolor{hellgrau} 133.6\,nm} & \multirow{-3}{1.2cm}{\centering 3.20\\$\pm$ 0.08} & \multirow{-3}{1.2cm}{\centering 1314.9\\$\pm$ 18.7} & \multirow{-3}{1.2cm}{\centering 8.6\\$\pm$ 0.1} & \multirow{-3}{1.2cm}{\centering 0.16\\$\pm$ 0.01} \\
		\hline  \hline
		\multicolumn{5}{|c|}{\cellcolor{dunkelgrau} Gold beam} \\
		\hline \hline &&&&\\ &&&&\\
		\multirow{-3}{1.8cm}{\centering \cellcolor{hellgrau} 129.1\,nm} & \multirow{-3}{1.2cm}{\centering 6.08\\$\pm$ 0.10} & \multirow{-3}{1.2cm}{\centering -} & \multirow{-3}{1.2cm}{\centering -} & \multirow{-3}{1.2cm}{\centering -} \\
		\hline
	\end{tabular}
	\caption{\label{tab::ergebnisse} \textit{Decay-time constants obtained from a fit (using eq. (\ref{eq::recofit})) of the time-resolved measurements close to the peak emission wavelength of the second excimer continuum (129.1\,nm and 128.6\,nm, respectively) and at its wings for the different ion beams. The intensity ratio $\frac{\text{I}_{\text{s}}}{\text{I}_{\text{t}}}$ of light from singlet decays to light from triplet decays is also given. In case of the gold beam the spectrum contained too little statistics to fit the long-lived component; in this case no values are quoted.}}
\end{table}

In addition, the intensity ratio

\begin{equation}
	\frac{I_s}{I_t} = \frac{\int \limits_{0}^{\infty} A_s \cdot exp\{-\frac{t}{\tau_s}\} \cdot dt}{\int \limits_{0}^{\infty} A_t \cdot exp\{-\frac{t}{\tau_t}\} \cdot dt} \label{eq::intratio}
\end{equation}

of light from singlet and triplet decays is given in table \ref{tab::ergebnisse}. This ratio was calculated without regarding light from recombination processes. Therefore, the results can best be compared with values presented in the literature where the data only have been fitted with a sum of two exponential functions, e.g. in refs. \cite{carvalho,hitachi_83}. Depending upon whether the light at intermediate times is taken as part of the fast or as part of the slow component of the light emission the ratio $\frac{\text{I}_{\text{s}}}{\text{I}_{\text{t}}}$ slightly changes, but by less than 1\%.

The general behaviour of the time structure of the scintillation light emitted by LAr (two time constants for singlet and triplet decays, respectively, and one time constant for delayed recombination; see eq. (\ref{eq::recofit})) was found to be the same all over the second excimer continuum; the emission and recombination time scales were found to be independent of the exact wavelength to first order. For the different exciting particles, however, differences were found: the light-emission time constants of the fast component for proton beam excitation are considerably shorter than those found for sulfur and gold beam excitation. 
The latter two are, however, in good agreement. The mechanism behind this behaviour could be the mixing process (see eq. (\ref{eq::mixing})) between singlet and triplet states in collisions with hot electrons. With the heavy ion beams a higher density of secondary electrons is produced than with protons due to their higher LET\footnote{Two-phase TPC detectors make also use of this fact, as a higher LET leads to a higher density of charged particles and, hence, to a higher probability for recombination. Thus, the scintillation signal is enhanced compared to the charge signal for particles with high LET (nuclear recoils, heavy ions) \cite{chepel}, which allows particle discrimination.}. 
Therefore mixing processes become more likely. In that case the fast component of the scintillation light emission would be composed of two parts: decays of excimer molecules in the singlet state which have been produced directly by the ion-beam pulse, and molecules in the singlet state which have been produced from molecules in the triplet state by mixing. The decay-time scale of the former component is simply governed by the life time of the singlet state, the decay-time scale of the latter component is additionally influenced by the time scale of the mixing process. 
In case the mixing-time constant is comparable or even slightly larger than the actual decay-time constant of excimer molecules in the singlet state the values obtained for $\tau_{\text{s}}$ from fitting the data would be only effective time constants, i.e. a convolution of both effects, rather than the pure life time of the singlet state. For similar time scales the two effects could not be resolved. 
In excitation of LAr with synchrotron radiation which had just enough energy to populate only the progenitor state of the excimer singlet indeed a singlet life time of (1.65\,$\pm$\,0.1)\,ns was found \cite{morikawa}. This could be the undisturbed life time of the singlet state, as in the experiments with synchrotron radiation the energy was too low to ionize argon atoms and produce free electrons. However, at present, alternative explanations cannot be excluded. In that sense, our values for $\tau_{\text{s}}$ represent only the time constant of the fast component of scintillation light emission rather than the life time of excimer molecules in the singlet state.

This possible explanation could also be checked by comparing other particles with different LET values. Table \ref{tab::liter} lists the fast decay time constants quoted in literature. Although the uncertainties of most studies are rather large compared to our values, and most of the values agree well within the error bars, a slight trend towards longer life times for particles with higher LET values ($\alpha$'s, neutrons, fission fragments) can be seen, for example in the studies by Hitachi et al. \cite{hitachi_83}. In contrast, the time constants published by Carvalho et al. \cite{carvalho} and Peiffer et al. \cite{peiffer} do not support our hypothesis. However, in the latter work it has been reported that the singlet-to-triplet intensity ratio rises with decreasing $\alpha$-energy, i.e. with rising LET. This observation could indeed be explained by the mixing process proposed above. In summary, it can be concluded that further studies are needed in order to support or rule out this possible explanation. 
Such studies could, for example, be performed in the gas phase with varying pressure and/or temperature, thereby varying the probability for collisions with hot electrons.

The life times of excimer molecules in the triplet state are on average slightly larger for proton beam excitation than those obtained with sulfur ions. The triplet life time is known to be influenced by chemical impurities \cite{heindl}, however, as was described above, the purity of our gas was at a high level, and, in particular, the same for the different ion beams.

Reduced triplet life times due to the Penning (excimer quenching) process,

\begin{equation}
	Ar_2^* + Ar_2^* \longrightarrow Ar_2^+ + 2Ar + e^-, \label{eq::penning} 
\end{equation}

as suggested in \cite{mei} would lead to a dependence of $\tau_{\text{t}}$ on the density of excimer molecules. If this happened within the track of one beam ion, a higher excitation density with sulfur ions could lead to the observed shorter triplet life times. A dependence on the beam current was, however, not observed in our experiments. This leads to the conclusion that there was no significant overlap of the tracks of the single beam ions. Bi-excitonic quenching, i.e.

\begin{equation}
	Ar^* + Ar^* \longrightarrow Ar^+ + Ar, \label{eq::biexitonic}
\end{equation}

only reduces the light yield \cite{doke_2}, but not the life time of the excimer molecules. Both quenching mechanisms are intrinsic features of LAr and capable of altering the light yield even in absence of any impurities. In principle, quen\-ching, if it occurred, should also affect the singlet life time $\tau_{\text{s}}$. However, for sulfur beam excitation, which yields the higher excitation density, a longer life time is observed than for protons. Therefore, we conclude that quenching is not the dominant process for the excitation densities studied in the present work, but the influence of the mixing process, eq. (\ref{eq::mixing}), on $\tau_{\text{s}}$ dominates. A deconvolution of the various effects is not possible in our data.

The recombination time constants, $\tau_{\text{rec}}$, are different but consistent for each beam species. Besides this, the time constants found in the present work are in tension with the recombination time constant reported in \cite{kubota_rec}, which is much shorter ((0.8\,$\pm$\,0.2)\,ns). This is at present not understood and should be addressed in future dedicated measurements.

The intensity ratio $\frac{\text{I}_{\text{s}}}{\text{I}_{\text{t}}}$ is clearly different for protons and sulfur ions as incident particles, independent of the beam current and the wavelength within the second excimer continuum. The proposed mechanism for this behaviour is again the mixing process (eq. (\ref{eq::mixing})), which is more pronounced for the heavier projectiles and increases the intensity of the fast decaying singlet component. However, also an enhanced production of progenitor states of excimer molecules in the singlet state with heavy ions could explain the singlet-to-triplet ratios observed, but it fails in explaining the different emission time constants of the fast scintillation light.
The parameter $\frac{\text{I}_{\text{s}}}{\text{I}_{\text{t}}}$ consequently allows particle discrimination between 10\,MeV-protons and low-energy heavy ions (however, still in the regime of electronic stopping) based on a tail-to-total pulse shape discrimination.
The huge difference between $\tau_{\text{s}}$ and $\tau_{\text{t}}$ in argon allows hereby low discrimination thresholds \cite{chepel}, a unique feature of LAr. The big difference in the fast and slow scintillation component for different particle beams can also be seen in fig. \ref{fig::psd_p_S}, where the intensity of the delayed component is clearly enhanced for proton-beam excitation.

However, the singlet-to-triplet ratio depends on the exact wavelength within the second excimer continuum. Table \ref{tab::ergebnisse} shows a clear trend towards a rising intensity ratio $\frac{\text{I}_{\text{s}}}{\text{I}_{\text{t}}}$ for shorter wavelengths. This is supposed to be due to the spectral splitting of the emission spectrum of singlet and triplet decays. The former peak at 126.6\,nm, the latter at 127.6\,nm (calculated from the data shown in \cite{morikawa}). Therefore, the singlet component is more dominant at shorter wavelengths. The general behaviour of the singlet-to-triplet ratio, however, also allows particle discrimination in wavelength-integrated measurements, which is especially important in terms of collecting as much scintillation light as possible. Pulse shape discrimination methods to distinguish between electrons and nuclear recoils in liquid argon have been demonstrated by several authors \cite{boulay_psd,peiffer,lippincott,pollmann,boulay_sim}.

\begin{figure}
	\centering
	\includegraphics[width=\columnwidth]{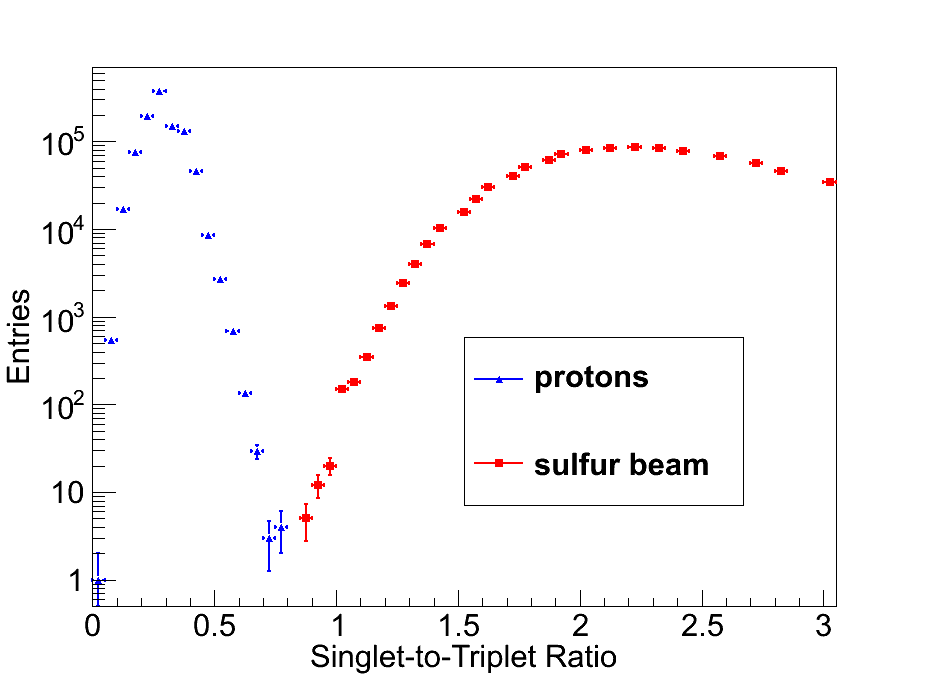}
	\caption{\label{fig::simulation} \textit{Simulated distributions of the singlet-to-triplet ratio with sulfur beam (squares; red in color online) and proton beam excitation (triangles; blue in color online) for a total number of 100 detected photons. Both distributions peak at the singlet-to-triplet ratios which were obtained in the time resolved measurements, 0.28 for protons and 2.19 for sulfur ions. For $\gtrapprox$\,100 detected photons the two distributions cease to overlap.}}
\end{figure}

To show how our results can be transferred to a detector system, a dedicated Monte-Carlo simulation was set up to simulate $\frac{\text{I}_{\text{s}}}{\text{I}_{\text{t}}}$ as it would be obtained in a measurement\footnote{Here and in the following we do not take into account any particular experimental aspects like optical coverage, as this simulation does not refer to a specific experiment. In contrast, we want to show the principle limits set by LAr itself.}. A certain number N of photons was generated; the probability P for each photon to be a singlet (triplet) photon was set to

\begin{equation}
	P_s = \frac{I_s}{I_s + I_t} = \frac{\frac{I_s}{I_t}}{1+\frac{I_s}{I_t}} \;\;\;\text{and}\;\;\; P_t = 1 - P_s ,
\end{equation}

where the values $\frac{\text{I}_{\text{s}}}{\text{I}_{\text{t}}} = 0.28$ and $\frac{\text{I}_{\text{s}}}{\text{I}_{\text{t}}} = 2.19$ were used for protons and sulfur ions, respectively. These are the intensity ratios obtained for the peak emission wavelength of the second excimer continuum, where most of the scintillation light is emitted. Figure \ref{fig::simulation} depicts the resulting singlet-to-triplet distributions for a total number of N\,=\,100 detected photons. 1,000,000 runs have been simulated. For a number of detected photons N\,$\gtrsim$\,100 the two distributions cease to overlap, hence, 100\% discrimination between incident protons with a kinetic energy of $\frac{\text{E}}{\text{m}} \approx 10\,\frac{\text{MeV}}{\text{amu}}$ and sulfur ions with $\frac{\text{E}}{\text{m}} \approx 3\,\frac{\text{MeV}}{\text{amu}}$ can be achieved. This is a principle limit set by the scintillation properties of LAr for the discrimination between these two species. For low-energy physics, of course, the 
discrimination between electrons and nuclear recoils is more important than the discrimination between protons and heavy ions, but as will be shown below, concerning the singlet-to-triplet intensity ratio protons behave much like electrons and sulfur ions like nuclear recoils, therefore, our results also provide some information for this case.

Taking a light yield of LAr of more than 40,000 photons per MeV, which was measured both with 17.8\,MeV protons and 1\,MeV electrons as incident particles and in absence of an external electric field \cite{doke88}, and assuming 100\% photon collection efficiency, 100 detected photons correspond to a deposited energy of only about 2.5\,keV (electron equivalent)\footnote{From the data presented here no information on the light yield of LAr with different incident heavy ion beams, i.e. information on quenching, can be derived, as no detector system absolutely calibrated in intensity has been used.}. This calculation of course disregards detector related effects like optical coverage or quantum efficiencies. For discrimination efficiencies lower than 100\% even lower thresholds are possible. A discrimination threshold of a few keV only is surely sufficient for $0\nu\beta\beta$-experi\-ments, likely also for direct Dark Matter search experiments, as well as for CNNS. Additional studies using incident particles 
with 
low energies have to be carried out to confirm this promising possibility.

In this simulation the following assumptions were made: Firstly, it was assumed that the values for $\frac{\text{I}_{\text{s}}}{\text{I}_{\text{t}}}$ given above are still valid at low energies deposited. As was mentioned in the beginning of sec. \ref{sec::results}, for heavy ions with kinetic energies smaller than $\sim$\,100\,$\frac{\text{keV}}{\text{amu}}$ nuclear stopping becomes the dominant process, and the stopping power becomes larger than would be expected from electronic stopping only \cite{northcliffe_schilling}. This might influence the mixing between molecules in the singlet and in the triplet state (eq. (\ref{eq::mixing})) and consequently change the ratio $\frac{\text{I}_{\text{s}}}{\text{I}_{\text{t}}}$ within the second excimer continuum.
This, however, has to be confirmed in wavelength- and time-resolved measurements at low energies deposited.
In fact, for $\alpha$-particles a dependence of $\frac{\text{I}_{\text{s}}}{\text{I}_{\text{t}}}$ on the particle's energy has been reported \cite{peiffer}. $\frac{\text{I}_{\text{s}}}{\text{I}_{\text{t}}}$ is rising for lower energies of the $\alpha$-particles.

Secondly, for the simulation the intensity ratios at the peak emission wavelength of the second excimer continuum have been taken. However, as can be seen from tab. \ref{tab::ergebnisse}, this ratio is not the same all over the continuum. Therefore, in a wavelength-integrated measurement the effective singlet-to-triplet ratio might slightly deviate from the values quoted above, thereby changing the discrimination threshold.

The time constants and singlet-to-triplet ratios found in the present work can be compared to the data quoted in literature (tab. \ref{tab::liter}). The time constants of the fast scintillation component obtained for the heavy ions sulfur and gold are found to be within the 1\,$\sigma$-error bars of nearly all the literature data, except the value given by Morikawa et al. \cite{morikawa}. As was discussed above, the latter value was obtained with synchrotron radiation and could therefore be the ''true'' life time of the molecules in singlet state. The fast emission time constants obtained with proton beam excitation are considerably shorter, but still of the same order of magnitude as the values quoted in literature. A problem in this comparison is that most of the literature data was obtained using a wavelength shifting material \cite{peiffer,lippincott,hitachi_83}, which induces a further uncertainty on the fast emission time scale. Only the data presented in \cite{heindl,heindl_epl,morikawa,carvalho} were 
measured directly.

The slow emission time scales presented here are similar to those quoted in literature, although there is quite a wide spread of these values, see table \ref{tab::liter}. As discussed above, chemical impurities influence the triplet life time, therefore, different purities of the LAr used might explain the differences. A major influence of the quenching effects discussed above can be excluded, as the spread in the slow emission time constants can also be found for the same exciting particle (see, e.g., $\alpha$-particles) with the same LET.

The singlet-to-triplet values we have obtained with proton beam excitation, especially those close to the peak emission wavelength of the second excimer continuum, are fully consistent with the ratios given in literature for particles with a low LET, electrons and $\gamma$'s. The $\frac{\text{I}_{\text{s}}}{\text{I}_{\text{t}}}$-ratio obtained with the sulfur beam, on the other hand, is of the same order as the literature values for heavily ionizing particles, $\alpha$'s, neutrons, and fission fragments, although once again a wide spread of these values is found. This shows that our studies with protons and sulfur ions as incident particles can help to understand the differences in the response of LAr to particles with different LET in general. These differences are used in low-energy physics to discriminate between electrons and nuclear recoils.
 
At longer wavelengths the time structure of the scintillation light of LAr was found to be partly different. The structure at $\sim$\,160\,nm, likely attributed to the analogue of the classical LTP in the liquid, could be fitted well by a convolution of a Gaussian and a single exponential function; see also fig. \ref{fig::gaussexpoconvolution}. An emission time constant of (39.1\,$\pm$\,2.7)\,ns was obtained.

The time structure of the spectral feature around 180 nm resembles the time structure of the second excimer continuum, and has two clearly distinguishable decay time constants (5.19\,$\pm$\,0.20\,ns and 1155.4\,$\pm$\,24.3\,ns for excitation with sulfur ions), which are similar to those obtained for the second excimer continuum. This hints to an energy transfer from the decay of vibrationally relaxed neutral argon excimer molecules to some faster light emitting process, possibly the second excimer continuum of xenon. The latter is found centering at 172\,nm in the gas phase \cite{morozov} and reported to be much faster than the second continuum emission of LAr \cite{morikawa}. An energy transfer is also possible, as the energy of Ar$_2^*$ is higher than the excitation energy of xenon \cite{efthimiopoulos,pollmann}. 
A dedicated measurement where the scintillation light of LAr was recorded wavelength-resolved with a pulsed beam, but only in a time interval of [40\,ns - 3180\,ns] after the trigger, showed that the light emission peaks at $\sim$\,174\,nm in that wavelength region. This could indeed be the second excimer continuum of xenon but redshifted by about 2\,nm in the surrounding argon matrix \cite{heindl_epl}.

For wavelengths longer than 190\,nm all light emissions were found to be very fast and only upper limits can be quoted due to the finite width of the beam pulse: $<$\,3.85\,ns at (201.6\,$\pm$\,1.5)\,nm and $<$\,4.15\,ns at (275.6\,$\pm$\,1.5)\,nm with sulfur-beam excitation, and $<$\,2.08\,ns at (210.6\,$\pm$\,3.0)\,nm and $<$\,2.04\,ns at (275.6\,$\pm$\,3.0)\,nm with proton-beam excitation. These emission features could be the third excimer continuum emitted in decays of highly charged ionic excimer molecules, which are found to be very fast already in the gas phase \cite{wieser}.

The total intensity of scintillation light emitted at wavelengths longer than 190\,nm is weak, however, this part of the spectrum could be strongly enhanced if the efficiency of the detector system rises steeply for longer wavelengths. In this case, the fast light emission of the third excimer continuum could mimic singlet photons from the second excimer continuum in wavelength-integrated measurements, thereby changing the $\frac{\text{I}_{\text{s}}}{\text{I}_{\text{t}}}$-ratio observed.

\begin{figure}
	\centering
	\includegraphics[width=\columnwidth]{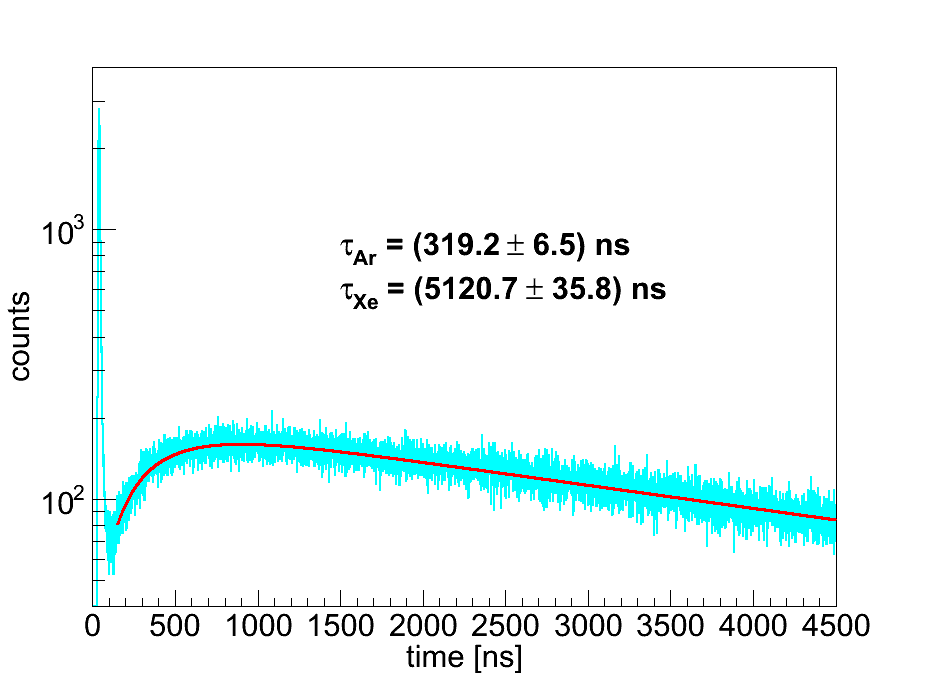}
	\caption{\label{fig::timestructure_xe} \textit{Time-resolved spectrum of LAr at 148.9\,nm (xenon impurity emission line) with sulfur beam excitation. The wavelength resolution is $\Delta\lambda$\,=\,1.5\,nm. The data was fitted with eq. (\ref{eq::xenonfit}) (solid line; red in color online). The light emission is built up with a time constant of (319.2\,$\pm$\,6.5)\,ns, the decay time constant is found to be (5120.7\,$\pm$\,35.8)\,ns. The spike at early times is light emission from the red wing of the second excimer continuum.}}
\end{figure}

The last spectral feature which has been investigated in a time-resolved measurement is the Xe-impurity emission line at 148.9\,nm. Its time structure shows a slow rise, followed by an even slower decay, see fig. \ref{fig::timestructure_xe}. This time structure could be fitted well with a sum of two exponential functions, 

\begin{equation}
	f_{xe}(t) = \frac{A}{\frac{\tau_{Ar}}{\tau_{Xe}}-1} \cdot \left(\exp\left\{-\frac{t-t_0}{\tau_{Ar}}\right\} - \exp\left\{-\frac{t-t_0}{\tau_{Xe}}\right\}\right) ,  \label{eq::xenonfit}
\end{equation}

which account for the rise with time constant $\tau_{\text{Ar}}$ and the decay with $\tau_{\text{Xe}}$. The time structure can be understood as an energy transfer from argon excimer molecules to the xenon impurity atoms, likely via long-range dipole-dipole coupling \cite{cheshnovsky}. Equation (\ref{eq::xenonfit}) is the analytical solution of the differential equations for the production and decay rates of xenon atoms in an excited state. For proton-beam excitation, where excimer molecules in the triplet state dominate, a time constant $\tau_{\text{Ar}}$\,=\,(1422.8\,$\pm$\,61.7)\,ns was found; for sulfur-beam excitation $\tau_{\text{Ar}}$\,=\,(319.2\,$\pm$\,6.5)\,ns. The time constants for the decay were found to be very similar, $\tau_{\text{Xe}}$\,=\,(5120.7\,$\pm$\,35.8)\,ns with sulfur ions and $\tau_{\text{Xe}}$\,=\,(5056.4 $\pm$\,116.5)\,ns with protons as exciting particles. 
These long decay-time constants could be explained if the xenon atoms are at first excited to the $^3\text{P}_1$ state, and subsequently transferred into the $^3\text{P}_2$ state by collisions \cite{cheshnovsky}. The latter, however, has no dipole allowed transitions to the ground state \cite{heindl} and radiates at 149.2\,nm \cite{nist}, which is close to the observed wavelength of 148.9\,nm. The radiative decay of xenon atoms in the $^3\text{P}_2$ state is in competition with the formation of xenon excimer molecules in collisions with xenon atoms in the ground state \cite{cheshnovsky}. These excimer molecules lead to the emission of the second excimer continuum of xenon which has indeed been observed in our data as mentioned above.

\section{Conclusions}
\label{sec::conclusion}

The scintillation light of LAr has been studied both in wavelength- and time-resolved measurements using protons, sulfur and gold ions as exciting particles. By far most of the scintillation light is emitted in the second excimer continuum, which is centered around $\sim$\,127\,nm. Light emission has been found to be very weak for all wavelengths longer than $\sim$\,170\,nm in the liquid phase. The spectral differences obtained for the different incident particles are tiny, therefore, particle discrimination using the spectral information only is not feasible.

In the present work, the time structure of the scintillation light was studied wavelength resolved for a set of certain wavelengths. The time structure of the emission of the second excimer continuum could be described by a sum of two exponential functions and a term accounting for recombination. The slow component, attributed to the decay of excimer molecules in the triplet state, is much more prominent for proton excitation than for heavier ions (sulfur). The intensity ratio of singlet-to-triplet decays consequently allows particle discrimination, which has already successfully been proven by several authors \cite{boulay_psd,peiffer,lippincott,pollmann,boulay_sim}. 
Under the assumption that the singlet-to-triplet intensity ratios obtained in this work, 0.28 for protons and 2.19 for sulfur ions, are still valid at low energies deposited, a discrimination threshold down to $\sim$\,2.5\,keV (electron equivalent, see section \ref{sec::timestudies}) seems possible. The latter value is obtained from a simulation assuming 100\% photon collection efficiency. 
The singlet-to-triplet ratios in the low-energy regime, however, have to be confirmed in studies with low energies of the incoming projectiles. Particle discrimination is feasible even in wavelength-integrated measurements, as the time structure of the second excimer continuum is the same all over its wavelength range, but the discrimination threshold might change slightly, as $\frac{\text{I}_{\text{s}}}{\text{I}_{\text{t}}}$ rises towards shorter wavelengths.
In experimental studies a discrimination threshold between electrons and nuclear recoils of a few keV only has already been demonstrated \cite{boulay_psd,boulay_sim}.

The decay time constants at longer wavelengths ($\lambda\,\gtrsim$ 190\,nm) were found to be very fast. In principle, this could disturb the determination of the singlet-to-triplet ratio in wavelength-integrated methods, as this fast light might mimic the singlet component of the second excimer continuum. The influence of the third excimer continuum on the intensity ratio would be biggest, if its intensity varied with the type of incident particle and if the detector sensitivity rose steeply for wavelengths above 190\,nm. The latter is often the case, see e.g. figure 5 in ref. \cite{heindl_epl}. However, our studies have shown that the third excimer continuum is strongly suppressed in LAr and very similar for all exciting particles, which helps to understand why wavelength-integrated particle identification works so well. 

An emission line at 148.9\,nm and a structure centered around $\sim$\,174\,nm could be identified to be due a residual xenon impurity.

\section{Acknowledgements}
\label{sec::danksagung}

The authors thank Dominikus Hellgartner for his help with RooFit. This work was supported by funds of the Maier-Leibnitz-Laboratorium (Garching) and the Excellence Cluster ''Origin and Structure of the Universe''.

\end{document}